\documentclass[twocolumn,prb,preprintnumbers,amsmath,amssymb]{revtex4}
\usepackage{graphicx}% Include figure files
\begin{document}
%\preprint{draft}

\title{Anomalous Temperature Dependence of Lower Critical Field in Ultraclean URu$_2$Si$_2$}

\author{Ryuji~Okazaki$^{1}$\thanks{E-mail address: okazaki@scphys.kyoto-u.ac.jp},  Masaaki~Shimozawa$^{1}$, Hiroaki~Shishido$^{1,2}$, Marcin~Konczykowski$^{3}$, Yoshinori~Haga$^{4}$, Tatsuma~D.~Matsuda$^{4}$, Etsuji~Yamamoto$^{4}$, Yoshichika~Onuki$^{4,5}$, Yoichi~Yanase$^6$, Takasada~Shibauchi$^{1}$, and Yuji~Matsuda$^{1}$}

\affiliation{
$^1$Department of Physics, Kyoto University, Kyoto 606-8502, Japan\\
$^2$Research Center for Low Temperature and Materials Sciences, Kyoto University, Kyoto 606-8501, Japan\\
$^3$Laboratoire des Solides Irradi\'es, CNRS-UMR 7642 \& CEA/DSM/DRECAM, Ecole Polytechnique, 91128, Palaiseau, France\\
$^4$Advanced Science Research Center, Japan Atomic Energy Agency, Tokai 319-1195, Japan\\
$^5$Graduate School of Science, Osaka University, Toyonaka, Osaka 560-0043, Japan\\
$^6$Department of Physics, Niigata University, Niigata 950-2181, Japan}

\begin{abstract}

To investigate a mysterious superconducting state of URu$_2$Si$_2$ embedded in the so-called hidden order state, the lower critical field $H_{c1}$ is precisely determined down to 55\,mK for  {\boldmath $H$}$\parallel a$ and  {\boldmath $H$}$\parallel c$.  For this purpose, the positional dependence of the local magnetic induction is measured on ultraclean single crystals ($T_c$ = 1.4\,K) with residual resistivity ratio exceeding 700.  We find that the temperature dependence of $H_{c1}$ significantly differs from that of any other superconductors.   The whole $H_{c1}(T)$ for {\boldmath $H$}$\parallel a$  are well explained by the two superconducting gap structures with line and point nodes, which have been suggested by the recent thermal conductivity and  specific heat measurements.   On the other hand, for {\boldmath $H$}$\parallel c$, a change of slope with a distinct kink in $H_{c1}(T)$, which cannot be accounted for by two gaps, is observed.   This  behavior for  {\boldmath $H$}$\parallel c$ sharply contrasts with the cusp behavior of $H_{c1}(T)$ associated with a transition into another superconducting phase found in UPt$_3$ and U$_{1-x}$Th$_x$Be$_{13}$.  The observed anomalous low-field diamagnetic response is possibly related to a peculiar vortex dynamics associated with chiral domains due to the multicomponent superconducting order parameter with broken time reversal symmetry.

\end{abstract}

\maketitle

\section{Introduction}

Exotic superconducting state with nontrivial Cooper pairing in heavy fermion systems continues to be a central focus of investigations in strongly correlated electron systems.\cite{Sig91,Pfl09}  Among them, URu$_2$Si$_2$ has mystified researchers since the heavy fermion superconductivity occurs deep inside the mysterious `hidden order' state whose transition temperature is $T_h$ = 17.5\,K.\cite{Pal85,Map86,Sch86}

Several salient features of the superconducting state have been reported in URu$_2$Si$_2$.   According to several experimental observations, most of the carriers ($\sim$\,90\,\%) disappear below $T_h$,\cite{Sch87,Bel04,Beh05,Kas07} resulting in a semimetallic electronic structure with a density one order of magnitude smaller than in other heavy fermion superconductors. Superconductivity with such a low density is remarkable since the superfluid density is very low in some way reminiscent of underdoped cuprates; the superconductivity by itself is an exceptional case of pairing among heavy electrons with a long Fermi-wavelength in a nearly semimetallic system. Moreover, pressure studies revealed that the superconductivity coexists with the hidden order phase having no intrinsic magnetic order but cannot coexist with the antiferromagnetic order with large magnetic moment,\cite{Uem05,Amitsuka,Has08} in contrast with other heavy fermion compounds where the superconductivity often coexists with magnetic orders.  Thus, although the genuine hidden order parameter is still an open question, the hidden order may provide an intriguing stage for a new type of unconventional superconducting state.

Despite these studies, very little is known about the exotic superconducting state of URu$_2$Si$_2$, mainly because the superconductivity is extremely sensitive to disorder.\cite{Ram91}  Recent studies using ultraclean single crystals with very large residual-resistivity-ratio ($RRR$) have enabled us to develop an advanced understanding about the superconducting state of URu$_2$Si$_2$.\cite{Kas07,Oka08}  A number of unprecedented superconducting properties have been reported, including superconductor-insulator-like first-order transition at upper critical field $H_{c2}$,\cite{Kas07} flux line lattice melting at sub-Kelvin temperatures \cite{Oka08} and quantum transport of quasiparticles.\cite{Ada08}  Moreover, it has been shown that below $T_h$ the nearly perfect compensation (equal number of electrons and holes) is realized.\cite{Kas07}  According to the quantum oscillation,\cite{Ohk99,Ber97,Shi09} electronic transport,\cite{Kas07} specific heat \cite{Yan08} and thermal conductivity measurements,\cite{Kas07} the presence of the light spherical hole band and anisotropic heavy electron band has been suggested.   This indicates that URu$_2$Si$_2$ is an essentially multiband superconductor.  Recent angle-resolved thermal conductivity and specific heat measurements have suggested two distinct superconducting gap structures having different nodal topology with horizontal line nodes in the hole band and point nodes in the electron band.\cite{Kas07,Yan08,Kas09,Mat06} From the group symmetry analysis, a chiral $d$-wave state with a form,
\begin{equation}
\Delta(\mbox{\boldmath {$k$}})=\Delta_0 k_z (k_x \pm ik_y),
\end{equation}
has been proposed, which breaks time reversal symmetry (TRS).\cite{Kas07,Yan08,Kas09}

Up to some years ago, the only known condensate with broken TRS was the $A$-phase of superfluid $^3$He.  More recently, several candidates of unconventional superconducting phases which break TRS have been reported in several classes of strongly correlated materials, including ruthenate Sr$_2$RuO$_4$,\cite{Luk98} filled skutterudite PrOs$_4$Sb$_{12}$,\cite{Aok03} and U-based heavy fermion compounds, UPt$_3$ \cite{Luk93} and U$_{1-x}$Th$_x$Be$_{13}$.\cite{Hef90}  In these materials,  muon spin relaxation measurements in zero field reveal the development of spontaneous magnetic moment below $T_c$ and in Sr$_2$RuO$_4$, nonzero Kerr rotation has been reported.\cite{Xia06} Theoretically, it is predicted  that in a superconductor with multicomponent order parameter, chiral domain walls separating different degenerate superconducting states are formed.\cite{Vol84,Sig89,Sig99}  Then the system should spontaneously generate supercurrents at the edge and the domain walls. Moreover, it is proposed that these chiral domains should strongly influence the vortex penetration especially at low fields.\cite{Ich05} These chiral domains and low-field magnetic response in unconventional superconductors with broken TRS have been explored by several experimental techniques,\cite{Mot02,Tam03,Cic05,Dol05,Kir07} however there is no direct evidence of such anomalies in the supercurrents as well as vortex penetration.  

In this paper, to shed further light on the exotic superconducting properties of URu$_2$Si$_2$, we investigated the low-field diamagnetic response of ultraclean single crystals.    The results provide strong evidence for the highly unusual superconducting state of URu$_2$Si$_2$ embedded in the hidden order state.

\section{Experimental}

High-quality single crystals of URu$_2$Si$_2$ with $RRR=700$ were grown by the Czochralski pulling method in a tetra-arc furnace.\cite{sample}  The well defined superconducting transition was confirmed by the specific heat measurements.  Experiments have been performed on single crystals with typical dimensions of $2.3\times0.75\times0.15$\,mm$^3$ (see the inset of Fig.~3) down to 55\,mK by using a dilution refrigerator.   

The standard technique to determine the lower critical field $H_{c1}$ is to measure the dc magnetization in the bulk crystals.   However, reliable determination of $H_{c1}$ is a difficult task in the presence of the flux pinning and magnetic relaxation.  To avoid these difficulties,  we determined $H_{c1}$ by directly detecting the positional dependence of the field $H_p$ at which flux penetration occurs from the edge of the crystal.\cite{Oka09}   To measure the local magnetic induction, we used micro Hall sensors tailored in a GaAs/AlGaAs heterostructure.\cite{Shi07}

First we studied the pinning properties in the crystals by measuring the magnetic field distribution in the vortex state, which is obtained by the scanning micro Hall sensor with an active area of $2\times2$\,$\mu$m$^2$.  The Hall sensor was scanned on the surface area of the crystal by using the piezoelectric device.   The distance between the Hall sensor and the crystal surface was kept constant by monitoring the tunneling currents.  Next we measured the local magnetic induction very precisely by using a miniature Hall-sensor array with each active area of $\sim 5\times 5$\,$\mu$m$^2$ (the center-to-center distance of neighboring sensors is 20\,$\mu$m).  The crystal was directly placed on top of the array (see the inset of Fig.~2(a)).

In both measurements, earth magnetic field was shielded by $\mu$-metal. The residual field at the sample space is less than 0.5\,$\mu$T.  In all measurements,  the external field was applied after the sample was zero-field cooled to the desired temperature from the temperature above $T_c$.

\section{Results}

\begin{figure}[t]
\begin{center}
\includegraphics[width=8.6cm]{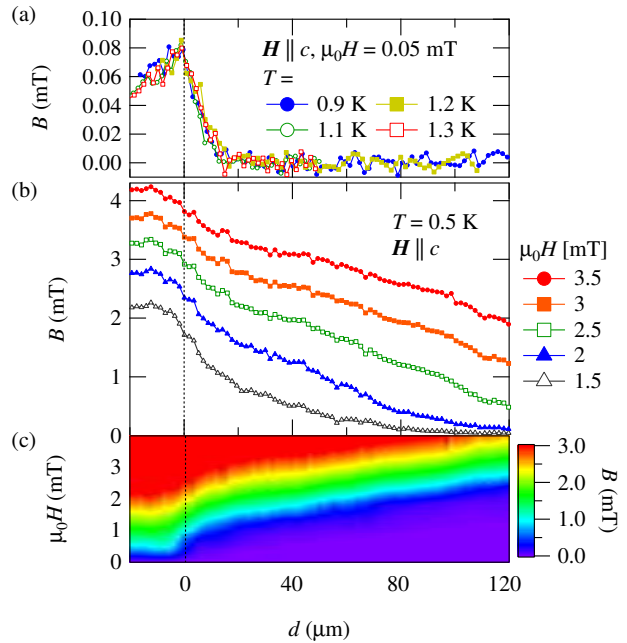}
\caption{(color online). The positional dependence of the local magnetic induction $B$ near the edge region of the crystal, which is determined by the scanning Hall-probe microscopy.  The dashed line is the position of the edge (For details, see the text.)   (a) The profile of magnetic induction in the Meissner state.  The external magnetic field ($\mu_0H$ = 0.05\,mT) is applied parallel to the $c$ axis.  (b) The profile of magnetic induction in the vortex state at several fields at $T$ = 0.5~K. (c) Contour image of (b).   }
\end{center}
\end{figure}

Figure~1(a) displays the profile of the magnetic induction near the edge of $ab$ plane in the Meissner state when the external magnetic field $\mu_0H$ = 0.05\,mT is applied parallel to the $c$ axis.  The profile is measured  by the scanning Hall-probe microscopy.  With approaching from the outside of the crystal, the magnetic induction is enhanced to $B$ $\sim$ 0.08\,mT followed by a sharp drop near the edge of the crystal.  The induction is perfectly screened inside the crystal.   Here we defined the edge at the peak of the local magnetic induction, as shown by the dotted line.  The enhancement of the magnetic induction at the edge is due to the field induced by the Meissner shielding currents flowing the edge region.\cite{Zel94}    The perfect screening is achieved in a short distance at $d\gtrsim 14$\,$\mu$m just below $T_c$,  where $d$ is the distance from the edge.  This demonstrates the homogeneous superconducting state of the crystal,

\begin{figure}[t]
\begin{center}
\includegraphics[width=8.6cm]{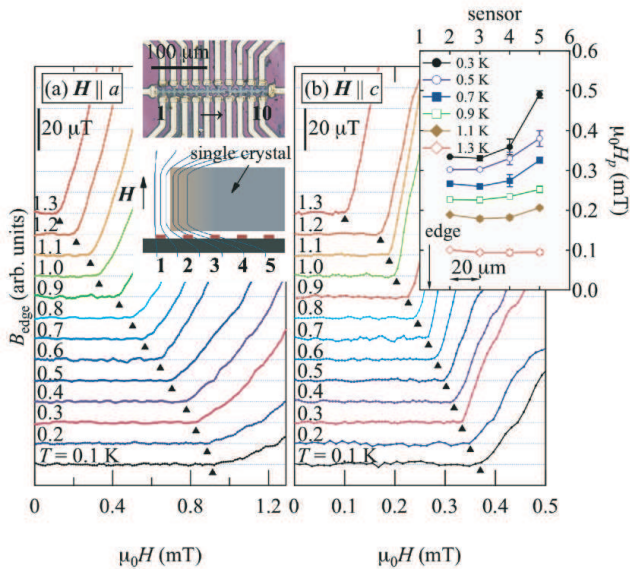}
\caption{(color online).  (a) Local magnetic induction $B_{\rm edge}$ measured by the sensor at the edge of the crystal (Sensor 2), as a function of $H$ for {\boldmath $H$}$\parallel a$. The triangles indicate the flux penetration field $H_p$. Each data is vertically shifted for clarity. Inset: Photograph of the miniature Hall-sensor array, and a schematic illustration of the experimental setup.  Sensor~1 locates just outside of the crystal.  (b) The same plot for {\boldmath $H$}$\parallel c$.  Inset is the positional dependence of $H_p$. }
\end{center}
\end{figure}

Figures~1(b) and (c) display the distribution of the local magnetic induction at several external magnetic fields associated with the flux penetration at $T = 0.5$\,K.   The magnetic induction decays  monotonically with the distance $d$ from the edge.  This indicates the Bean critical state dominated by the bulk flux pinning. In this case, $H_{c1}$ is determined from the local magnetic induction near the edge, at which the first flux penetration is most sensitively detected.\cite{Oka09} This is in contrast to the case when the geometrical surface barrier (determined by sample shape) is important rather than bulk pinning, in which the magnetic flux distribution shows a dome-like shape and the flux penetration field can be detected at the center of the crystal.\cite{Shi07}

The positional dependence of the local magnetic induction $B_{\rm edge}$ were measured accurately by using the Hall-sensor array near the edge for {\boldmath $H$}$\parallel a$ and {\boldmath $H$}$\parallel c$.   The sample edge is located between the Hall sensors 1 and 2 (see the inset of Fig.~2), which is confirmed by the Meissner response of each sensor, as shown in Fig.~1(a).  In fact, sensor 2 exhibits a perfect Meissner response, while slight enhancement of $B$ from $\mu_0H$ is observed in sensor 1.  Figures\:2(a) and (b) display the field dependence of $B_{\rm edge}$ measured by sensor 2.  The flux penetration fields $H_p$ shown by the triangles are clearly resolved by the deviation from the Meissner state ($B_{\rm edge}=0$).   The inset of Fig.\:2(b) depicts the positional dependence of $H_p$.  Obviously $H_p$ is position-independent at sensors 2 and 3  which locate close to the edge ($d\leq40\,\mu$m) even at low temperatures.  On the other hand, $H_p$ is enhanced at sensors 4 and 5, which locate at $d>40$\,$\mu$m,  at low temperatures.  These results again represent the Bean critical state in this crystal and assure that  we can accurately determine $H_{c1}$ from $H_p$ measured by sensors near the edge.

\begin{figure}[t]
\begin{center}
\includegraphics[width=8.6cm]{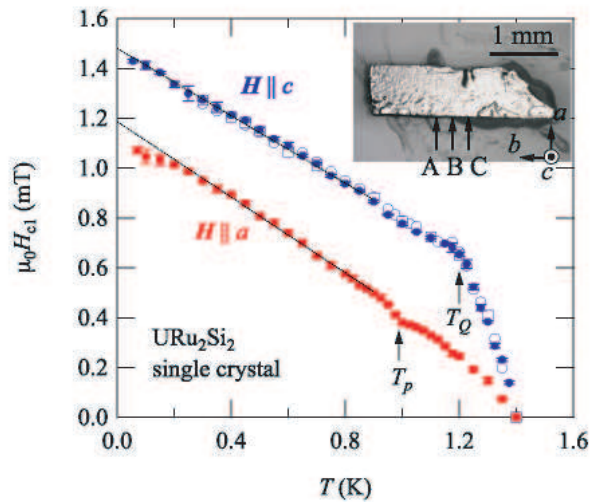}
\caption{(color online). Inset: Photograph of the single crystal URu$_2$Si$_2$ used in the present study. Local magnetic induction is measured at the edge regions A, B, and C. Main panel: Temperature dependence of the lower critical fields $H_{c1}$ for {\boldmath $H$}$\parallel a$ (solid squares) and {\boldmath $H$}$\parallel c$ (solid circles), which are determined at the edge region B shown in the inset. Arrows indicate a kink anomaly at $T_Q$ = 1.2\,K and a cusp behavior at $T_p$ $\simeq$ 1\,K. The dotted lines are the fits to the linear dependence. Open circles and open squares show $H_{c1}^c(T)$ determined at the edge regions A and C shown in the inset, respectively.}
\end{center}
\end{figure}

Figure\:3 depicts the temperature dependence of $H_{c1}^a$ and $H_{c1}^c$, where $H_{c1}^a$  and $H_{c1}^{c}$ are the lower critical fields for {\boldmath $H$}$\parallel a$ ({\boldmath $H$}$\parallel c$), respectively.  We evaluate $H_{c1}^c=3.82 H_p$ and $H_{c1}^{a}=1.15 H_p$  by using the expression 
\begin{equation}
H_{c1}=\frac{H_p}{\tanh \sqrt{0.36b/a}},
\end{equation}
where $a$ and $b$ are the width and the thickness of the crystal, respectively, for a platelet sample.\cite{Bra99}
This expression takes into account the geometrical effect in an nonellipsoidal shape, where the demagnetization effect in ideal ellipsoids is modified.\cite{Bra99}
The magnetic penetration depth is evaluated from $H_{c1}$ to $\lambda_{a} \simeq 0.8$ $\mu$m, which is quantitatively consistent with the $\mu$SR results of $\lambda_{a} = 0.7 - 1$ $\mu$m.\cite{Knetsch,Amato} 
According to the Ginzburg-Landau theory, the temperature dependence of $H_{c1}$ anisotropy, $\gamma_{H_{c1}} \equiv H_{c1}^{c}/H_{c1}^{a}$, may be different from that of $H_{c2}$ in multiband superconductors.\cite{Kogan,Mir03} At $T\to T_c$, however, $\gamma_{H_{c1}}$ should coincide with the $H_{c2}$ anisotropy. 
The experimentally determined $\gamma_{H_{c1}}\simeq 3$ close to $T_{c}$ is then fully consistent with the reported $H_{c2}$ anisotropy near $T_c$,\cite{Oka08} giving us confidence in the accuracy of the result.
As seen in Fig.~3, $H_{c1}^a$ increases linearly below $\sim 0.8$~K with decreasing temperature.  A cusp structure can be seen at $T_p=$1~K.   Meanwhile, $H_{c1}^c$ increases steeply below $T_c$ and increases linearly after exhibiting a distinct change in the slope with a kink at $T_Q=1.2$~K.  As shown by the dotted lines, $H_{c1}^c$ at low temperatures  increases linearly down to 55\,mK, while $H_{c1}^a$ exhibits a tendency to saturation below 200\,mK.

\section{Discussion}

First we discuss the $T$-dependence of $H_{c1}$ in the low temperature regime.  Since $H_{c1}$ is proportional to the superfluid density, the $T$-linear dependence of $H_{c1}^c$ down to low temperatures indicates the presence of line nodes in the superconducting gap function.  Moreover, the tendency to saturation in $H_{c1}^a(T)$ below 200\,mK indicates that the line nodes are located parallel to the $ab$ planes (horizontal node).  This is because  the supercurrents always flow parallel to the nodal planes for {\boldmath $H$}$\parallel c$, while for {\boldmath $H$}$\parallel a$ supercurrents have a component which flows across the horizontal nodes. This component is perpendicular to the velocities of the quasiparticles around the nodes, which reduces the contributions of nodal quasiparticles to the superfluid density.  Since in the multiband superconductors the penetration depth and hence the lower critical field are governed by the band with light mass,  it is natural to consider that the horizontal line node locates in the spherical light hole band.  This is  consistent with the previous results of angle-resolved thermal conductivity.\cite{Kas09,Mat06}

Most remarkable features are anomalies at $T_Q$ for {\boldmath $H$}$\parallel c$ and $T_p$ for {\boldmath $H$}$\parallel a$.   In the specific heat measurements, no anomaly has been observed at $T_p$ in the present crystals with large $RRR$ values.  However, a peak structure in the heat capacity has been reported in the vicinity of $T_p$ for crystals with $RRR<$ 100, possibly due to the inhomogeneous distribution of $T_c$.\cite{sample}  Therefore we cannot rule out a possibility that the observed $T_p$-anomaly in $H_{c1}^a$ is due to a tiny portion with low-$T_c$ phase in the crystal. On the other hand, no anomaly has been reported in the heat capacity at $T_Q$ for any crystals, indicating no evidence of the low-$T_c$ phase as an origin for the $T_Q$-anomaly.  It is also highly unlikely that the anomaly at $T_Q$ arises from the inhomogeneous flux penetration of the crystal, because of the following reasons.   First, the open squares and circles in Fig.~3 show $H_{c1}^c$ determined at different edge regions in the same crystal.  The distinct kink anomaly at $T_Q$ is quite well reproduced.  Second, as shown in Fig.~1(a), the Meissner shielding currents flow homogeneously and the crystal has well defined edge.  These results lead us to conclude that the anomaly of $H_{c1}^c(T)$ at $T_Q$ is intrinsic. We stress that the anomaly at $T_Q$ has never been reported in the crystals with lower $RRR$ values.\cite{Wuc93}

\begin{figure}[t]
\begin{center}
\includegraphics[width=8.6cm]{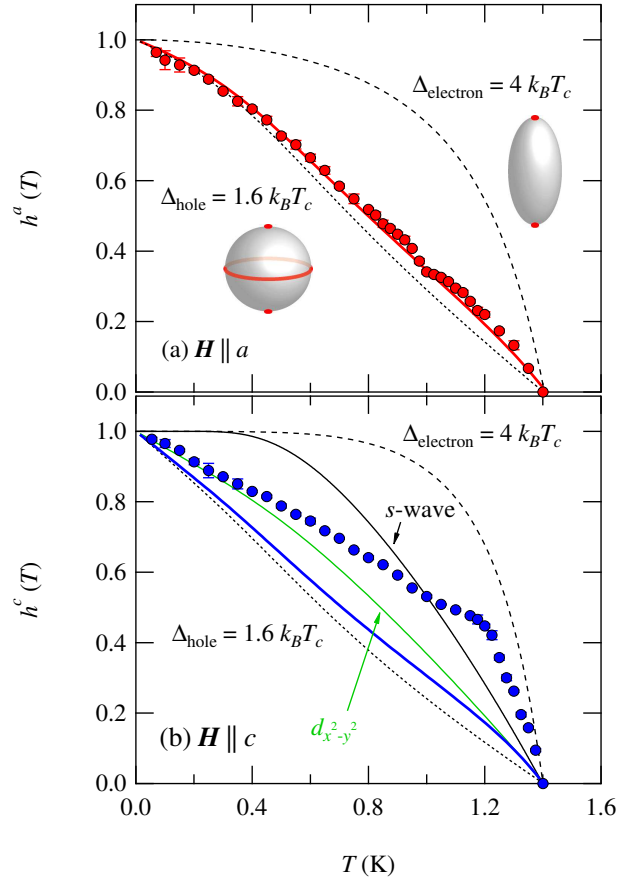}
\caption{(color online). Comparison of $h^{a,c}\equiv H_{c1}^{a,c}(T)/H_{c1}^{a,c}(0)$ (symbols) with (a) $\sqrt{n_s^a n_s^c}$ and (b) $n_s^a$, respectively. The dotted and dashed lines indicate the superfluid densities for the hole band with $\Delta_h = 1.6k_BT_c$ and the electron band with $\Delta_e = 4.0k_BT_c$, respectively, which are obtained by assuming the illustrated nodal topologies suggested by the thermal conductivity measurements.\cite{Kas07} The solid thick and thin lines represent the results of the two-gap fitting and the calculations for other gap symmetries, respectively.}
\end{center}
\end{figure}

Now we try to fit the temperature dependence of $H_{c1}^a$ and $H_{c1}^c$ in accordance with the multiband model.  Recent angle-resolved thermal conductivity and specific heat measurements revealed two distinct superconducting gap structures having different nodal topology with horizontal line nodes in the hole band and point nodes in the electron.\cite{Kas07,Yan08,Kas09}  To see whether multiband effect can explain anomalous behaviors of the lower critical fields, we discuss the temperature dependence of $H_{c1}^a$ and $H_{c1}^c$ in terms of the two-gap model below.  In the two-band model, the in-plane and out-of-plane superfluid density normalized by their values at $T=0$\,K, $n_s^a$ and $n_s^c$, respectively, can be written as
\begin{equation}
n_s^i(T) = x^in^i_{h}(T)+(1-x^i)n^i_{e}(T),\quad i=a, c
\end{equation}
where $n_h^i$ and $n_e^i$ are the normalized superfluid density of hole and electron bands, respectively. $x^i$ is the ratio of the electron and hole contributions given by
\begin{equation}
x^i = \frac{X^i_h}{X^i_h+X^i_e},\quad X^i_{h(e)} = \int\frac{(v_{F,{h(e)}}^i)^2}{|\mbox{\boldmath $v$}_{F,{h(e)}}|}dS_{h(e)},
\end{equation}
where $v_{F,h(e)}^i$ is the Fermi velocity of holes (electrons).\cite{Prozorov}  The hole mass determined by the dHvA measurements is nearly isotropic $m_h^a\approx m_h^c\approx 13m_0$,\cite{Ohk99,Shi09} where $m_h^a$ $(m_h^c)$ is the mass of the hole along the $a$ $(c)$ axis and $m_0$ is the free electron mass.  The heavy electron mass is anisotropic and is estimated to be $m_e^a\approx 85 m_0$ and $m_e^c \approx 305 m_0$ from the large Sommerfeld coefficient in the heat capacity ($\gamma \sim80$\,mJ/K$^2$mol) \cite{Map86} and the anisotropy of upper critical field $H_{c2}$.\cite{Oka08}  Then we obtain $x^a\approx$ 0.87 and $x^c \approx$ 0.95.  Since theses values are close to unity, the superfluid density, particularly its temperature dependence at low temperatures, is governed by the hole band.  This again provides a strong support to the horizontal line nodes located on the hole bands as mentioned before.  Moreover, the presence of point nodes has been suggested along the $c$ axis in the heavy electron bands.\cite{Kas07,Yan08}   Then, we calculate the superfluid density by assuming the gap functions $\Delta_h(\mbox{\boldmath $k$},T)=\Delta_h(T)\times 2 \sin \theta \cos \theta$ with point and horizontal line nodes for hole bands and $\Delta_e(\mbox{\boldmath $k$},T)=\Delta_e(T) \times \sin \theta$ with $c$-axis point nodes for electron bands.

The normalized lower critical fields,  $h^{a,c}(T)\equiv H_{c1}^{a,c}(T)/H_{c1}^{a,c}(0)$, are related to the superfluid density as $h^c=n_s^a$ and $h^a=\sqrt{n_s^a n_s^c}$.  First we try to fit $h^a(T)$.  Because the hole band dominates at low temperatures,  $\Delta_h(\mbox{\boldmath $k$},T)$ can be determined by $h^a(T)$.  The best fit is obtained by $\Delta_h(0)=1.6k_BT_c$ and $\Delta_e(0)=4.0k_BT_c$.  As shown by the thick line in Fig.\:4(a), the fitting result reproduces well the overall temperature dependence of $h^a(T)$, including the observed tendency toward saturation at low temperatures. We note that this $\Delta_h(0)$ value is close to $\Delta_h(0)\sim\hbar v_{F,h}/\pi \xi_h \sim 1.5k_BT_c$, which is obtained from $v_{F,h}\sim2\times10^4$\,m/s \cite{Ohk99} and $\xi_h\sim25$\,nm.  Here $\xi_h=\sqrt{\Phi_0/2 \pi \mu_0 H_{c2}^h}$ is the coherence length of the hole band estimated from the ``virtual upper critical field" of the hole band $\mu_0H^h_{c2} \sim 0.5$\,T.\cite{Kas07}  In Fig.\:4(b),  $h^c(T)$ calculated by using the same $\Delta_h(0)$ and $\Delta_e(0)$ values is plotted by the thick line.  In sharp contrast to $h^a(T)$, the calculation strongly deviates from the data.  We tried to fit the data by assuming various $\Delta_h(0)$ and $\Delta_e(0)$ values and other gap symmetries, but could not reproduce the data, particularly the anomaly at $T_Q$.

Based on these results,  we conclude that the multigap effect cannot be an origin of the anomaly at $T_Q$.  Although a change in the slope of $H_{c1}(T)$ has been reported in UPt$_3$ \cite{UPt3} and (U$_{1-x}$Th$_{x}$)Be$_{13}$,\cite{Hef90,UBe13} there are crucial differences from our observation.  First, in URu$_2$Si$_2$ $H_{c1}(T)$ is strongly suppressed below $T_Q$, while in UPt$_3$ and (U$_{1-x}$Th$_{x}$)Be$_{13}$ $H_{c1}(T)$ is enhanced below the kink temperature.  Second, in URu$_2$Si$_2$ the kink anomaly is observed solely for {\boldmath $H$}$\parallel c$, while in UPt$_3$ and (U$_{1-x}$Th$_{x}$)Be$_{13}$ the anomaly is observed in any field directions. In the latter compounds, the kink anomaly has been attributed to the transition into another superconducting phase with different gap structure.  

The observed strong suppression of $H_{c1}^c(T)$ indicates the drastic change of the vortex entry to the sample at $T_Q$.  Here we try to explain qualitatively this anomalous behavior of $H_{c1}(T)$  in terms of a peculiar vortex dynamics associated with chiral domains due to the multicomponent superconducting order parameter with broken TRS. It is natural to consider that the superconducting order parameter couples to the hidden order, because the superconductivity appears only in the hidden order phase as shown by recent pressure studies.\cite{Amitsuka} To discuss the pairing state in the presence of the hidden order, we investigate the dimensionless Ginzburg-Landau free energy which describes the pairing state with  $k_z(k_x+ik_y)$ form at zero magnetic field and low temperature, 
\begin{equation}
\begin{split}
f &= \frac{T-T_{\rm c}}{T_{\rm c}} |\eta_{1}|^{2} 
  + \frac{T-T'_{\rm c}}{T_{\rm c}} |\eta_{2}|^{2} 
  + \frac{(|\eta_{1}|^{2} + |\eta_{2}|^{2})^{2}}{2} \\
& + \beta_2 \frac{(\eta_{1}\eta_{2}^{*} -\eta_{1}^{*}\eta_{2})^{2}}{2}
  + \beta_3  |\eta_{1}|^{2} |\eta_{2}|^{2}. 
\end{split}
\end{equation}
Here the order parameter is written as 
\begin{equation}
\Delta_{e,h}(\mbox{\boldmath $k$}) = \eta_{1} \phi_{e,h}^{\rm xz}(\mbox{\boldmath $k$}) + \eta_{2} \phi_{e,h}^{\rm yz}(\mbox{\boldmath $k$}), 
\end{equation}
where $\phi_{e,h}^{\rm xz}(\mbox{\boldmath $k$})$ and $\phi_{e,h}^{\rm yz}(\mbox{\boldmath $k$})$ are the order parameters having $k_zk_x$ and $k_zk_y$ symmetry, respectively. 
When the system has a four-fold rotation symmetry in the $ab$ plane, 
these pairing states must have the same transition temperature, and 
$T'_{\rm c}=T_{\rm c}$. 
Then, the TRS is spontaneously broken just below $T_{\rm c}$ within the 
BCS theory since $\beta_2 > 0$ and $\beta_3 - 2 \beta_2 < 0$. 
One the other hand, the four-fold rotation symmetry may be broken by the 
hidden order which occurs at $T_h$ = 17.5~K. 
When this is true, the transition temperature of the $k_zk_x$ state 
differs from that of the $k_zk_y$ state, 
namely $T'_{\rm c} \ne T_{\rm c}$. 
(We can assume $T'_{\rm c} < T_{\rm c}$ without loss of the generality.) 
Then, the pairing state without broken TRS 
$(\eta_{1}, \eta_{2}) \propto (1,0)$ is stable just below $T_{\rm c}$ 
and the second-order phase transition to the pairing state 
$(\eta_{1}, \eta_{2}) \propto (1, \pm {\rm i} \alpha)$ occurs at 
\begin{equation}
T_{\rm c2} = T_{\rm c} - \frac{T_{\rm c} - T'_{\rm c}}{2 \beta_{2} - \beta_{3}} < T_{\rm c}. 
\end{equation}
The TRS is broken at this second superconducting transition. 

Assuming $T_{\rm c2}=T_Q$, the order parameter with broken TRS can produce domain walls along the $c$ axis at $T_Q$. As suggested in Ref.~30, magnetic fields then can penetrate inside through the domain wall even in the Meissner state.  
Thus the formation of domain walls gives rise to a peculiar reduction of the penetration field only for {\boldmath $H$} $\parallel c$.  As shown in Figs.~1(a), (b) and (c), our scanning Hall probe detects no apparent inhomogeneous field distribution, which implies that such domains, if exist, are smaller than the size of Hall sensor ($\sim2\,\mu$m).  We also note that the kink anomaly at $T_Q$ has never been observed in the crystals with low $RRR$ values,\cite{Wuc93} implying that the domain formation is sensitive to the impurities. It should be noted that a possible presence of such a domain structure, which influences the vortex dynamics, has been suggested in UPt$_3$, U$_{1-x}$Th$_{x}$Be$_{13}$, Sr$_2$RuO$_4$ and PrOs$_4$Sb$_{12}$.\cite{Mot02,Cic05}

The second superconducting transition discussed here is similar to the transition from the $p$-wave pairing state with TRS to the chiral $p$-wave state without TRS which has been proposed for Sr$_2$RuO$_4$.\cite{Agterberg} The four-fold rotation symmetry is broken by the in-plane magnetic field in Sr$_2$RuO$_4$, while the hidden order may spontaneously break the four-fold rotation symmetry in URu$_2$Si$_2$ without applying any external field. We note that local four-fold symmetry breaking in the hidden order phase has been recently proposed theoretically.\cite{Har10}

\section{Conclusion}

We have determined accurately the lower critical fields of ultraclean URu$_2$Si$_2$  by the positional dependence of the local magnetic induction.   The lower critical fields  exhibit several distinct features, which have never been observed in any other superconductors.  The temperature dependence of $H_{c1}$ at low temperatures suggests the horizontal line nodes in the energy gap in the light hole band.  We show that the whole $H_{c1}(T)$ for {\boldmath $H$}$\parallel a$ can be explained by the two gaps with line and point nodes, which is consistent with the previous reports.  In sharp contrast, for {\boldmath $H$}$\parallel c$ we find a distinct kink in the slope of $H_{c1}(T)$ at 1.2\,K, which cannot be accounted for by the two gaps.  This anomalous low-field diamagnetic response has been discussed in the light of the multicomponent order parameter with broken TRS.  The highly unusual superconducting state of URu$_2$Si$_2$  deserves further theoretical and experimental investigations.

\section*{Acknowledgement}

We thank D.F.~Agterberg, E.H.~Brandt, H.~Ikeda, K.~Machida and M.~Sigrist for helpful discussions and V.~Mosser for providing Hall sensors. This work was supported by Grant-in-Aid for the Global COE program ``The Next Generation of Physics, Spun from  Universality and Emergence", Grant-in-Aid for Scientific Research on Innovative Areas ``Heavy Electrons'' (No. 20102006) from MEXT of Japan, and KAKENHI from JSPS.

\end{document}